\documentclass[aps,prd,floatfix,superscriptaddress,preprintnumbers,floatfix]{revtex4} 
\usepackage{amssymb}
\usepackage{amsmath}
\usepackage{amsfonts}
\usepackage{orcidlink}
\usepackage{epsfig}             
\usepackage{graphicx}
\usepackage{stackrel}
\usepackage{tabularx}
\usepackage{color}
\usepackage{dsfont}
\usepackage[T1]{fontenc}

\definecolor{darkerblue}{rgb}{0.0,0.0,0.5}
\newcommand{\seq}{\begin{subequations}}
\newcommand{\sen}{\end{subequations}}

\newcommand{\eq}{\begin{eqnarray}}
\newcommand{\en}{\end{eqnarray}}

\begin{document}

\title{Charge-exchange reactions with pion and kaon beams \\
  in the NA64h experiment at CERN} 

\author{Sergei N.~Gninenko \orcidlink{0000-0001-6495-7619}} 
\affiliation{Millennium Institute for Subatomic Physics at
the High-Energy Frontier (SAPHIR) of ANID, \\
Fern\'andez Concha 700, Santiago, Chile}
\affiliation{Institute for Nuclear Research, 117312 Moscow, Russia}
\affiliation{Joint Institute for Nuclear Research 141980 Dubna, Russia}

\author{Sergey~Kuleshov~\orcidlink{0000-0002-3065-326X}}
\affiliation{Millennium Institute for Subatomic Physics at
the High-Energy Frontier (SAPHIR) of ANID, \\
Fern\'andez Concha 700, Santiago, Chile}
\affiliation{Center for Theoretical and Experimental Particle Physics,
Facultad de Ciencias Exactas, Universidad Andres Bello,
Fernandez Concha 700, Santiago, Chile}

\author{Valery~E.~Lyubovitskij \orcidlink{0000-0001-7467-572X}}
\affiliation{Institut f\"ur Theoretische Physik, Universit\"at T\"ubingen, \\
Kepler Center for Astro and Particle Physics, \\ 
Auf der Morgenstelle 14, D-72076 T\"ubingen, Germany} 
\affiliation{Millennium Institute for Subatomic Physics at
the High-Energy Frontier (SAPHIR) of ANID, \\
Fern\'andez Concha 700, Santiago, Chile}
\affiliation{Department of Physics, Tomsk State University, 634050 Tomsk, 
Russia}

\author{Alexey~S.~Zhevlakov \orcidlink{0000-0002-7775-5917}} 
\affiliation{Bogoliubov Laboratory of Theoretical Physics,
JINR, 141980 Dubna, Russia} 
\affiliation{Matrosov Institute for System Dynamics and 
  Control Theory SB RAS, \\  Lermontov street, 134,
  664033, Irkutsk, Russia} 

\begin{abstract}

We discuss a new approach to search for dark sector coupled to quarks
in the invisible decays and oscillations of light neutral mesons produced
in the charge-exchange (CEX) reactions with pion and kaon beams
$\pi^-(K^-)+ (A, Z) \to \pi^0, \eta, \eta' (\bar K^0) + (A,Z-1)$
at nuclei target $(A,Z)$ in the NA64h experiment at CERN for $P_{\rm Lab}$
running in the interval from 5 to 50 GeV. For estimating of the projection
sensitivity for the proposed searches the knowledge of the meson yield
is crucial. This work is dedicated to the accurate evaluation of CEX cross
sections for the wide range of energies and target nuclei which could be
also useful for other experiments.

\end{abstract}

\maketitle
	
\section{Introduction}

\par In this paper we discuss the perspectives of a novel experiment
to search for dark sector particles coupled to quarks in the invisible decays
and oscillations of light neutral mesons produced  in the charge-exchange (CEX) 
reactions of negative  pion and kaon beams
$\pi^-(K^-)+ (A, Z) \to \pi^0, \eta (\bar K^0) + (A,Z-1)$ at nuclei target
$(A,Z)$ in the NA64h experiment at CERN for $P_{\rm Lab}$ running in the interval
from 5 to 50 GeV~\cite{NA64h:2024mah}-\cite{Gninenko:2025xmb}.
The first results reported by NA64h from a proof-of-concept search for dark sectors
via invisible decays of pseudoscalar $\eta$  and $\eta'$ mesons
at the CERN SPS~\cite{NA64h:2024mah} looks very promising, demonstrating
the great potential of the new approach. They provide clear guidance on how to extend
and enhance the sensitivity of future searches for dark sector with  invisible neutral
meson decays, see, e.g., recent Ref.~\cite{Gninenko:2025xmb}, pointing, however,
at the same time at a lack and needs of improving our knowledge of CEX cross sections
in particular for the range of energies and target nuclei planned to used
by NA64h~\cite{NA64h:2024mah}.  
CEX reactions is a powerful and unique tool to create energetic neutral
mesons to study their invisible modes in the NA64h experiment. We need to understand
CEX reactions cross sections at high-accuracy level to perform comprehensive analysis
of decays modes opening a window into New Physics.
CEX reactions have good signature (energy and angle distribution)
for study of rare effects.

\par Results of experimental studies of the CEX reactions on nuclei
with pion beams $\pi^- + (A, Z) \to M^0 + (A,Z-1)$ at nuclei target
$(A,Z)$ with $M^0 = \pi^0$, $\eta$, $\eta'$, $\omega$, $f_2(1270)$
have been reported in Refs.~\cite{Barnes:1976ek,Apokin:1981iv}.
Study of the CEX reactions with kaon beams has been performed
in Refs.~\cite{Astbury:1966ufw}. 
To clarify an importance of the kaon beams in study of hadron structure
one should also mention the planned $K$-long facility (KLF) experiment
at JLab~\cite{Strakovsky:2016azh}
based on colliding of the $K_L$ mesons at proton target. The main
objective of the KLF experiment is the study of the structure of
hyperons and pentaquarks in the reactions $K_L p$ producing two and three
hadrons in the final state.

The objective of the present paper is unified description of the CEX
reactions using the Regge formalism~\cite{Regge}.
Detailed overview of the progress reached in study of the CEX reactions
can be found in Ref.~\cite{Gninenko:2023rbf}.
Also in Ref.~\cite{Gninenko:2023rbf} we consider
application of our formalism to the CEX reactions it charged pion beams
has been done in Ref.~\cite{Gninenko:2023rbf}.
Obtained cross section have been used in analysis of invisible decays of
light neutral pseudoscalar mesons~\cite{NA64h:2024mah}.
For global analysis of the CEX reactions see Ref.~\cite{Nys:2018vck}. 
In present paper we follow the strategy: 

(1) We study the integral cross sections
of the CEX reactions at proton targe $\pi^- + p \to \pi^0 + n$, 
$\pi^- + p \to \eta [\to 2 \gamma]  + n$, $\pi^- + p \to \eta' [\to 2 \gamma] + n$, and
$K^- + p \to \bar K^0 + n$ based on data and existing formulas
discussed in literature (see Refs.~\cite{Barnes:1976ek,Apokin:1981iv} 
and~\cite{Astbury:1966ufw}).
Fitting formulas for integral cross sections are based
the Regge formalism valid for $s \ge s_0 = 10$ GeV$^2$.

(2) We extend our analysis to CEX reactions on arbitrary target nuclei.
 Extension to nuclei $N=(A,Z)$ with charge $Z$ is normally
done by multiplying by a factor $Z^{2/3}$. However, we found
that this  should be corrected (as
$Z^{2/3-0.15 Z^{-2/3}}$, see below) which leads to some deviation
from the behavior of $\sim Z^{2/3}$. In our analysis
we use data for the CEX reactions at $P_{\rm Lab} = $ 40 GeV~\cite{Apokin:1981iv}. 
We focus on the production of the $\pi^0$, $\eta$, and $\eta'$.

(3) Using our knowledge on the $\pi^0$, $\eta$, and  $\eta'$ 
production in the CER reactions at arbitrary nuclei target
we extend our formalism to the reaction with kaon beam
$K^- + (A,Z) \to \bar K^0 + (A,Z-1)$. 

(4) We are careful in estimate of errors. We found that
the data on the integral cross sections of the CEX
reactions $\pi^- + p \to \pi^0 + n$, 
$\pi^- + p \to \eta(\eta') [\to 2 \gamma] + n$, and
$K^- + p \to \bar K^0 + n$ cross section
are fitted with accuracy of 10\%. 

The manuscript is organized as follows.
In Section II we discuss our formalism and apply it for description
of the CEX with pion and kaon beams. In Section III we present our
conclusions. 

\section{CEX reactions with pion and kaon beams in Regge formalism}

For many decades the most convenient formalism for the description of
the hadronic collisions at large $s$ and small $t$ with $s \gg |t|$
is the Regge formalism~\cite{Regge}. 
It is based on the idea that the corresponding cross section scales as
$\Big(\dfrac{s}{s_0}\Big)^{2 \alpha - 2}$, where $\alpha$ is the intercept
of the Regge trajectory, $s$ is the total energy squared and $s_0 = 10$ GeV$^2$
is its initial value. Further details can be found in Ref.~\cite{Gninenko:2023rbf}.
In particular, a generic expression for the integral cross section of
the CEX reaction is parametrized by two parameters $\sigma_0$ and $\alpha$ as
\eq
\sigma(s) &=& \sigma_0 \, \Big(\frac{s}{s_0}\Big)^{2 \alpha - 2} \,. 
\en
The parameters $\sigma_0 = \sigma(s_0)$ and $\alpha$ are fitted from data.
In Ref.~\cite{Gninenko:2023rbf} we focused on the region of high values of $s$.
Here we consider the huge interval of $s$ starting from $s_0$ till very high energies
(few hundreds of GeV$^2$). Our idea to include small values of $s$ in the fit is based
on perspectives to perform the experiment on the CEX reactions at CERN PS
where the $P_{\rm Lab}$ runs from 5 to 15 GeV, which corresponds to $s$ running
in the interval $10-30$ GeV$^2$ in the case of proton target. 

From our analysis we found that the data on the integral cross section of the
CEX reactions 
$\pi^- + p \to \pi^0 + n$, $\pi^- + p \to \eta [\to 2 \gamma] + n$, 
$\pi^- + p \to \eta' [\to 2 \gamma] + n$, and
$K^- + p \to \bar K^0 + n$ can be reproduced by the Regge formalism with a high accuracy.
The final expressions for the fitted integral cross sections including errors $\sim 10\%$
read 
\eq
\sigma_H^{\pi^-}(s) &=& (109 \pm 13) \,
\Big(\frac{s}{s_0}\Big)^{2 \alpha_{\pi} - 2} \, \mu{\rm b}
\,, \nonumber\\
\sigma_H^{\eta}(s) &=& (29 \pm 2) \,
\Big(\frac{s}{s_0}\Big)^{2 \alpha_{\eta} - 2} \, \mu{\rm b}
\,, \nonumber\\
\sigma_H^{\eta'}(s) &=& (1.05 \pm 0.10) \,
\Big(\frac{s}{s_0}\Big)^{2 \alpha_{\eta'} - 2} \, \mu{\rm b}
\,, \nonumber\\
\sigma_H^{\bar K^0}(s) &=& (145 \pm 25) 
\, \Big(\frac{s}{s_0}\Big)^{2 \alpha_{K} - 2} \, \mu{\rm b}
\,,
\en
where
\eq
\alpha_{\pi} = 0.405,
\qquad \alpha_{\eta} = 0.260,
\qquad \alpha_{\eta'} = 0.200,
\qquad \alpha_{K^0} = 0.248 
\,.
\en
Our predictions for the $P_{\rm Lab} = $ 50 GeV/s are
\eq
\sigma_H^{\pi^-} = (8.09 \pm 0.38) \, \mu{\rm b}\,, \qquad
\sigma_H^{\eta}  = (1.04 \pm 0.07) \, \mu{\rm b}\,, \qquad
\sigma_H^{\eta'}  = (0.029 \pm 0.003) \, \mu{\rm b}\,, \qquad
\sigma_H^{\bar K^0} = (4.91 \pm 0.68) \, \mu{\rm b}\,. 
\en
Extension to arbitrary nuclei is to multiply with factor $Z^{2/3 - 0.15/Z^{2/3}}$ 
\eq\label{sigma_fit}
\sigma_{Z} = \sigma_H  \, Z^{2/3 - 0.15/Z^{2/3}} \,. 
\en
Note in Ref.~\cite{Gninenko:2023rbf} we found that
the above formula is slightly modified. In particular,
we modified the power of the nuclei charge $Z$ as
$Z^{2/3} \to Z^{2/3 - 0.15/Z^{2/3}}$ to describe data
for heavier nuclei. 

\begin{figure}[t]
\begin{center}

  \includegraphics[scale=.80]{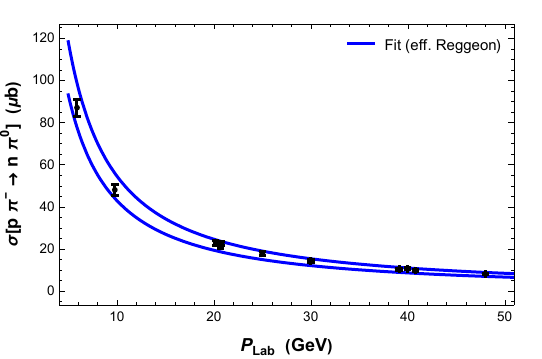}
  \hspace*{.2cm}
  \includegraphics[scale=.80]{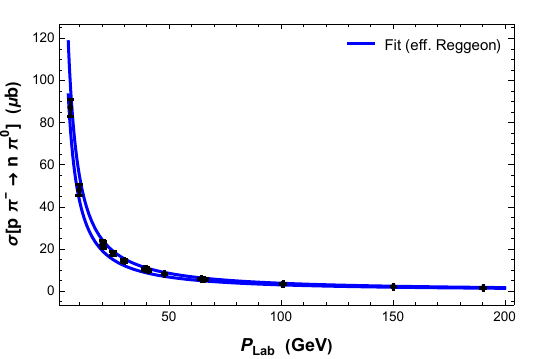}
  
\end{center}
\caption{Fit of the integral cross section of the $\pi^0$ production
  $\pi^-  + p \to \pi^0 + n$ for different intervals of the $P_{\rm Lab}$. 
    \label{fig:1}}

\begin{center}
  \includegraphics[scale=.80]{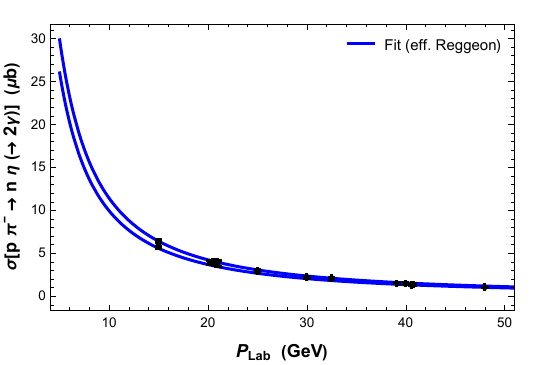}
  \hspace*{.2cm}
  \includegraphics[scale=.80]{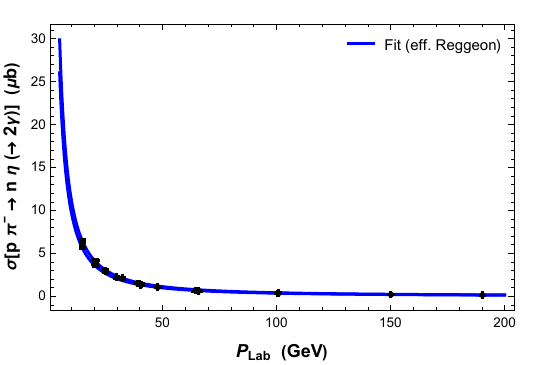}

\end{center}
\caption{Fit of the integral cross section of the $\eta$ production
  $\pi^-  + p \to \eta [\to 2 \gamma] + n$ for different intervals
  of the $P_{\rm Lab}$. 
    \label{fig:2}}

\begin{center}
  \includegraphics[scale=.80]{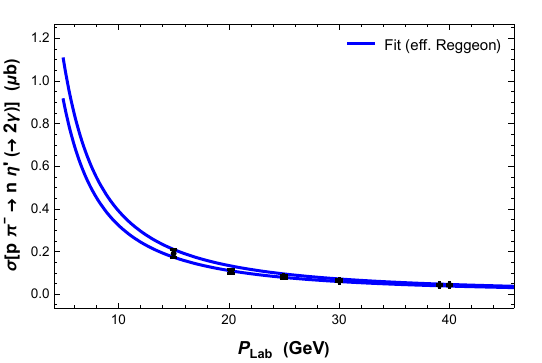}
  \hspace*{.2cm}
  \includegraphics[scale=.80]{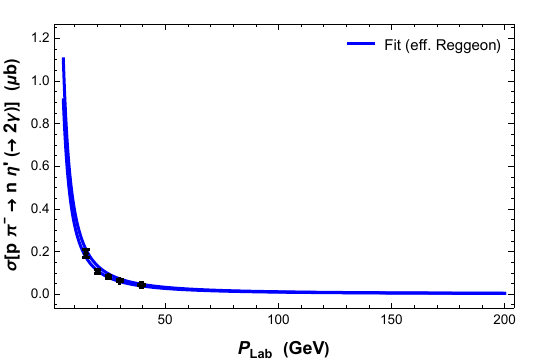}
\end{center}
\caption{Fit of the integral cross section of the $\eta'$ production
  $\pi^-  + p \to \eta' [\to 2 \gamma] + n$ for different intervals
  of the $P_{\rm Lab}$.
      \label{fig:3}}

\begin{center}
  \includegraphics[scale=.80]{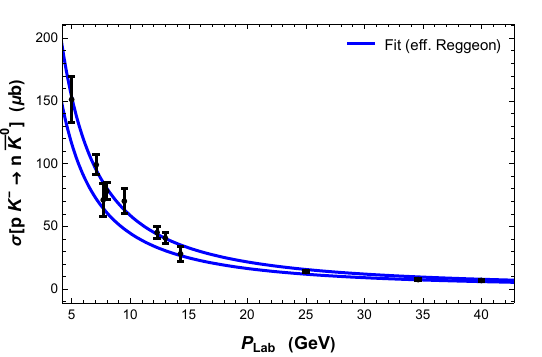}
  \hspace*{.2cm}
  \includegraphics[scale=.80]{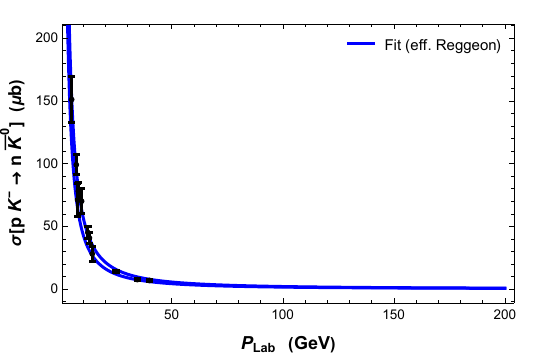}
\end{center}
\caption{Fit of the integral cross section of the $\bar K^0$ production
  $K^-  + p \to \bar K^0 + n$ for different intervals of the $P_{\rm Lab}$.
      \label{fig:4}}
\end{figure}

In Tables~\ref{tab:1}-\ref{tab:4} 
we make the comparison of the
theoretical fit and data for the integral cross sections
of the charge-exchange reactions
$\pi^-  + p \to \pi^0 + n$,
$\pi^-  + p \to \eta [\to 2 \gamma] + n$,
$\pi^-  + p \to \eta' [\to 2 \gamma] + n$, 
and $K^-  + p \to \bar K^0 + n$, 
at beam momentum $P_{\rm Lab}$ running up to 200 GeV.
Also for completeness we present there the relative error
between central values of theory and data $R = \dfrac{\sigma_{H, {\rm th}}^{\pi^0}
-\sigma_{H, {\rm data}}^{\pi^0}}{\sigma_{H, {\rm data}}^{\pi^0}}$ in $\% $.  
In Tables~\ref{tab:5} and \ref{tab:6} we present similar analysis
for the CEX reactions on different nuclei for the case of $\pi^0$
and $\eta$ production. 
In Figs.~\ref{fig:1}-\ref{fig:4} we show our results and comparison
with available data for the production of $\pi^0$, $\eta$, and $\bar K^0$
in charge exchange reactions at proton target at different values of
$P_{\rm Lab}$ running up to 200 GeV and at different nuclei.

\begin{figure}[t]
\begin{center}
  \includegraphics[scale=.80]{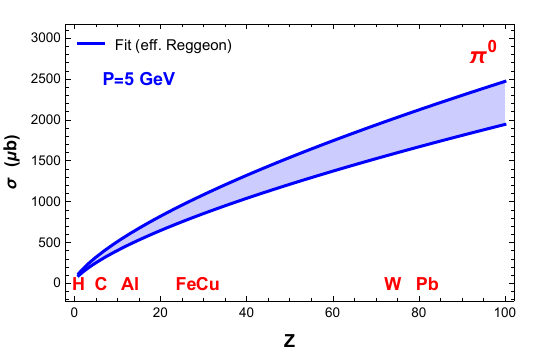}
  \hspace*{.2cm}
  \includegraphics[scale=.80]{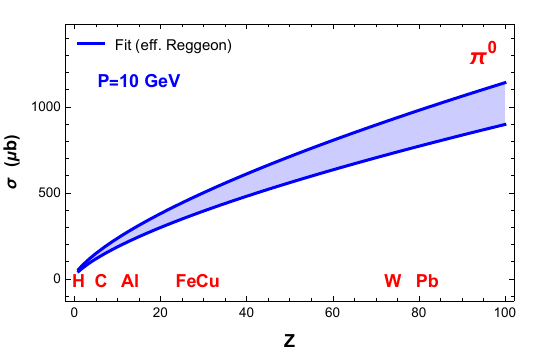}

  \includegraphics[scale=.80]{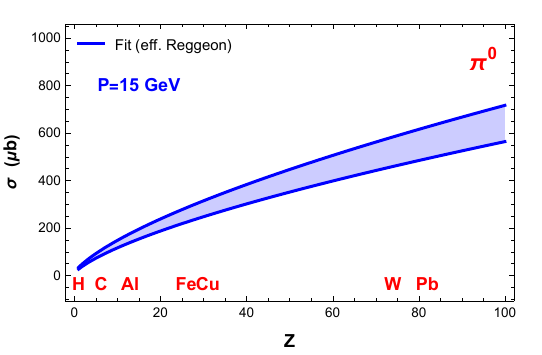}
  \hspace*{.2cm}
  \includegraphics[scale=.80]{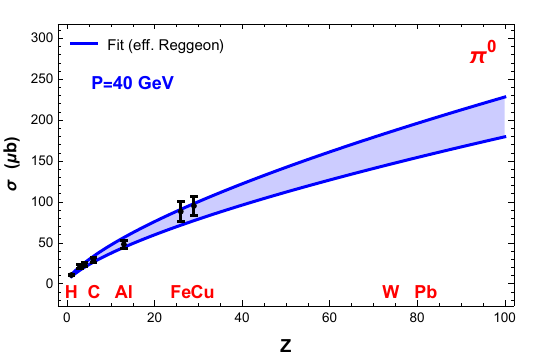}

\end{center}
\caption{Results for the integral cross section of the $\pi^0$ production
  at nuclei $\pi^-  + (A,Z) \to \pi^0 + (A,Z-1)$ for different values
  of the $P_{\rm Lab}$. 
    \label{fig:5}}
\end{figure}

\begin{figure}[t]
\begin{center}
  \includegraphics[scale=.80]{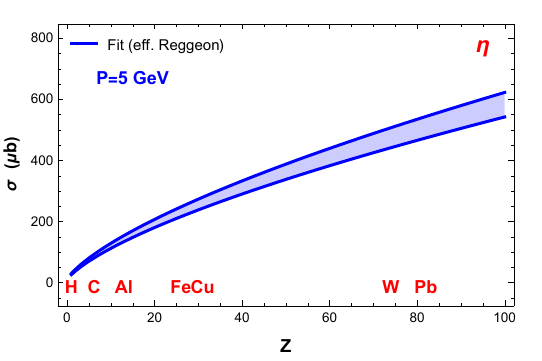}
  \hspace*{.2cm}
  \includegraphics[scale=.80]{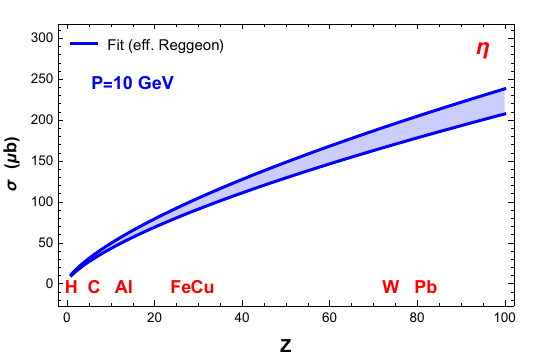}

  \includegraphics[scale=.80]{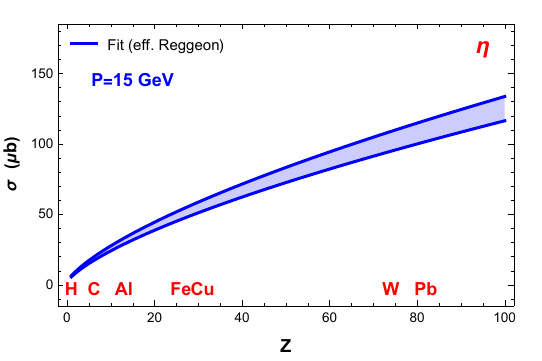}
  \hspace*{.2cm}
  \includegraphics[scale=.80]{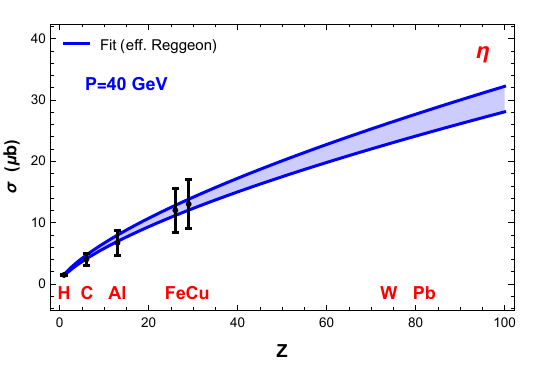}
\end{center}
\caption{Results for the integral cross section of the $\eta$ production
  at nuclei $\pi^-  + (A,Z) \to \eta [\to 2 \gamma] + (A,Z-1)$
  for different values of the $P_{\rm Lab}$.
  \label{fig:6}}
\end{figure}

\begin{figure}[t]
\begin{center}
  \includegraphics[scale=.80]{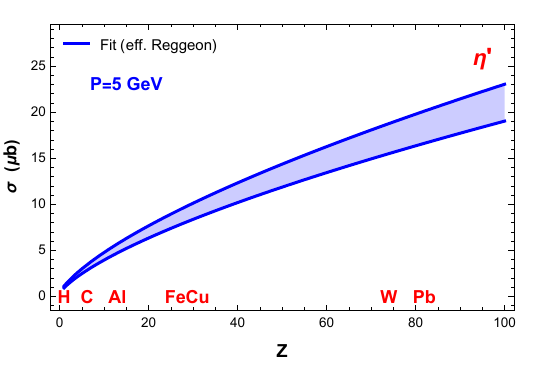}
  \hspace*{.2cm}
  \includegraphics[scale=.80]{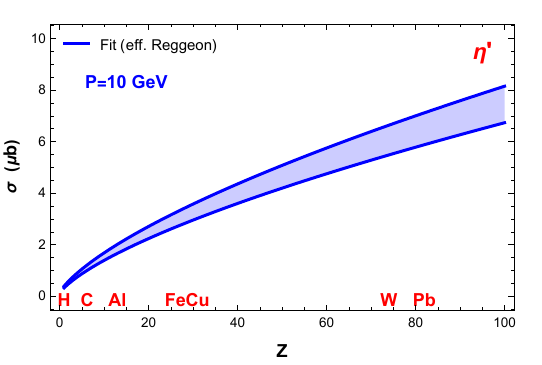}

  \includegraphics[scale=.80]{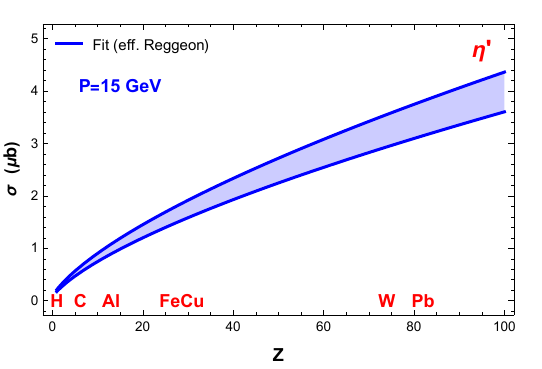}
  \hspace*{.2cm}
  \includegraphics[scale=.80]{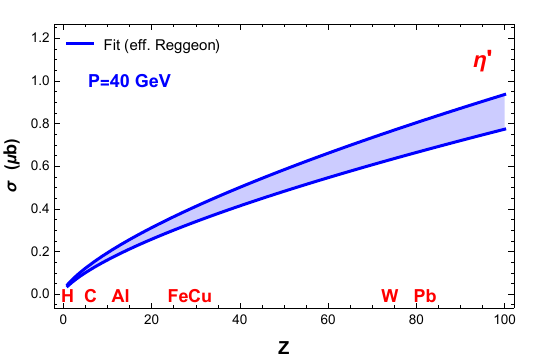}

\end{center}
\caption{Results for the integral cross section of the $\eta'$ production
  at nuclei $\pi^-  + (A,Z) \to \eta' [\to 2 \gamma] + (A,Z-1)$
  for different values of the $P_{\rm Lab}$.
      \label{fig:7}}
\end{figure}

\begin{figure}[t]
\begin{center}
  \includegraphics[scale=.80]{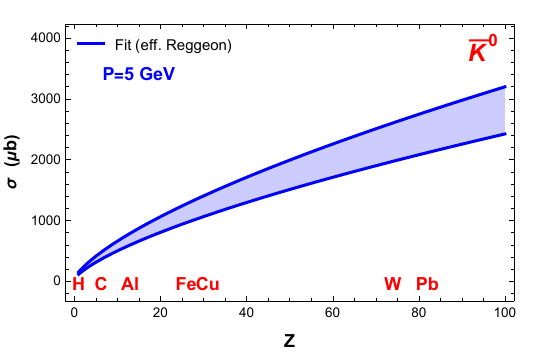}
  \hspace*{.2cm}
  \includegraphics[scale=.80]{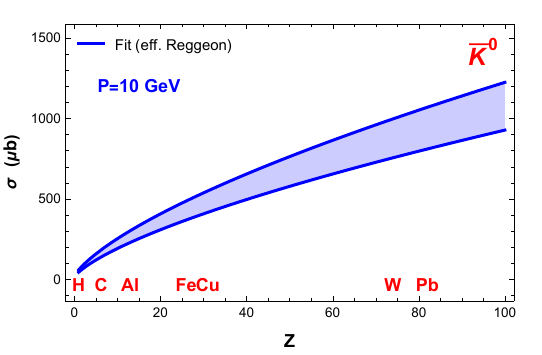}

  \includegraphics[scale=.80]{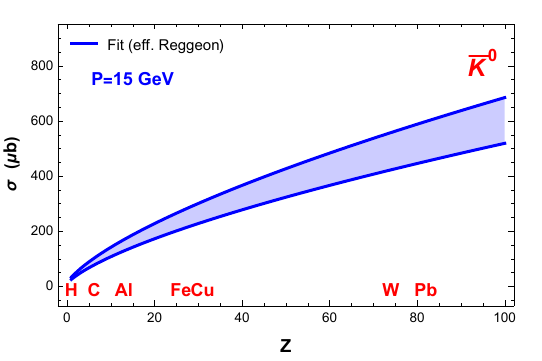}
  \hspace*{.2cm}
  \includegraphics[scale=.80]{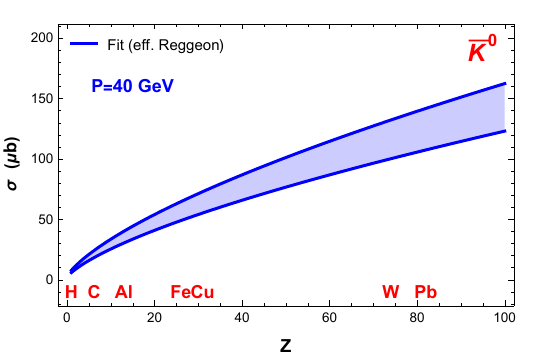}

\end{center}
\caption{Results for the integral cross section of the $\bar K^0$ production
  at nuclei $K^-  + (A,Z) \to \bar K^0 + (A,Z-1)$
  for different values of the $P_{\rm Lab}$.
  \label{fig:8}}
\end{figure}

Finally, in Figs.~\ref{fig:5}-\ref{fig:8} present the results for
the different values of the $P_{\rm Lab}$  
for the CEX reactions
$\pi^-  + (A,Z) \to \pi^0 + (A,Z-1)$,
$\pi^-  + (A,Z) \to \eta [\to 2 \gamma] + (A,Z-1)$,
$\pi^-  + (A,Z) \to \eta' [\to 2 \gamma] + (A,Z-1)$,
$K-  + (A,Z) \to \bar K^0 + (A,Z-1)$ at the following set of the nuclei
$Z$ = H, Li, W, Pb, H, Li, Be, C, Al, Fe, Cu, W, Pb including $10\%$ errors. 

The corresponding formulas for the integral cross sections for all
CEX modes read: 
\eq
\sigma_Z^{M}(s) = \sigma_H^{M}(s)  Z^{2/3 - 0.15/Z^{2/3}} 
\,.
\en

\begin{table}[ht!] 
\begin{center}
  \caption{Comparison of the theoretical fit and data
    for the integral cross sections of the charge-exchange reaction
    $\pi^-  + p \to \pi^0 + n$
     at beam momentum $P_{\rm Lab}$ running up to 200 GeV.}

\vspace*{.1cm}

\def\arraystretch{1.2}
\hspace*{-1cm}
\begin{tabular}{|c|c|c|c|c|c|c|c|c|c|}
\hline
\hline
$P_{\rm Lab}$ (GeV) & 5 & 9.8 & 13.3 & 15 & 18.2 & 20.2 & 20.8 & 25 & 30 \\
\hline
$\sigma_{H, {\rm data}}^{\pi^0}$ ($\mu{\rm b}$)
& $ 87 \pm 4$
& $ 48 \pm 2.5$
& $ 36 \pm 2$
& $ 34.2 \pm 1.8$
& $ 24   \pm 2.5$
& $ 22.8 \pm 1.1$
& $ 22.6 \pm 1.1$
& $ 17.8 \pm 0.8$
& $ 14.3 \pm 0.7$
\\
\hline
$\sigma_{H, {\rm th}}^{\pi^0}$ ($\mu{\rm b}$) 
& $ 88.16 \pm 10.51$
& $ 49.94 \pm 5.96$
& $ 35.22 \pm 4.20$
& $ 30.67 \pm 3.66$
& $ 24.52 \pm 2.92$
& $ 21.72 \pm 2.59$
& $ 21.00 \pm 2.50$
& $ 16.94 \pm 2.02$
& $ 13.69 \pm 1.63$
\\
\hline
$R$ (in \%) & 
1.3 & 4.0 & $-$ 2.2 & $-$ 1.0 & 2.2 & $-$ 4.7 & $-$ 7.1 & $-$ 4.8 & $-$ 4.3 \\
\hline
\hline
$P_{\rm Lab}$ (GeV) & 39.1 & 40 & 40.8 & 48 & 64.4 & 66 & 100.7 & 150.2 & 190.3 \\
\hline
$\sigma_{H, {\rm data}}^{\pi}$ ($\mu{\rm b}$)
& $ 10.4 \pm 0.5$
& $ 10.52 \pm 0.5$
& $ 9.8 \pm 0.3$
& $ 8.1 \pm 0.28$
& $ 5.61 \pm 0.25$
& $ 5.5 \pm 0.30$
& $ 3.20 \pm 0.14$
& $ 2.02 \pm 0.09$
& $ 1.44 \pm 0.06$ \\
\hline
$\sigma_{H, {\rm th}}^{\pi^0}$ ($\mu{\rm b}$) 
& $ 10.03 \pm 1.20$
& $  9.76 \pm 1.16$
& $  9.54 \pm 1.14$
& $  7.88 \pm 0.94$
& $  5.57 \pm 0.66$
& $  5.41 \pm 0.65$
& $  3.28 \pm 0.39$
& $  2.04 \pm 0.24$
& $  1.54 \pm 0.18$
\\
\hline
$R$ (in \%)
& $-$ 3.6
& $-$ 7.2
& $-$ 2.7
& $-$ 2.7
& $-$ 0.7
& $-$ 1.6
&   2.6 
&   1.1
&   7.1
\\
\hline
\end{tabular}
\label{tab:1}
\end{center}

\begin{center}
  \caption{Comparison of the theoretical fit and data
    for the integral cross sections of the charge-exchange reaction
    $\pi^-  + p \to \eta [\to 2 \gamma] + n$
     at beam momentum $P_{\rm Lab}$ running up to 200 GeV.}

\def\arraystretch{1.2}
\begin{tabular}{|c|c|c|c|c|c|c|c|c|}
\hline
\hline
$P_{\rm Lab}$ (GeV) & 15 & 20.2 & 20.7 & 20.8 & 21 & 25 & 30 & 32.5 \\
\hline
$\sigma_{H, {\rm data}}^{\eta; 2 \gamma}$ ($\mu{\rm b}$)
& $ 6.20 \pm 0.4$
& $ 3.90 \pm 0.2$
& $ 4.00 \pm 0.3$
& $ 3.85 \pm 0.24$
& $ 3.80 \pm 0.30$
& $ 2.92 \pm 0.15$
& $ 2.22 \pm 0.13$
& $ 2.10 \pm 0.10$
\\
\hline
$\sigma_{H, {\rm th}}^{\eta; 2 \gamma}$ ($\mu{\rm b}$) 
& $ 6.00 \pm 0.41$
& $ 3.90 \pm 0.27$
& $ 3.77 \pm 0.26$
& $ 3.74 \pm 0.26$
& $ 3.69 \pm 0.25$
& $ 2.86 \pm 0.20$
& $ 2.20 \pm 0.15$
& $ 1.95 \pm 0.13$
\\
\hline
$R$ (in \%) & 
- 3.4 & 0.02 & - 5.9 & - 2.9 & -3.0 & -1.9 & -1.1 & - 1.1 \\
\hline
\hline
$P_{\rm Lab}$ (GeV) & 39.1 & 40 & 40.8 & 48 & 64.4 & 101 & 150.2 & 190.3 \\
\hline
$\sigma_{H, {\rm data}}^{\eta; 2 \gamma}$ ($\mu{\rm b}$) 
& $1.50 \pm 0.04$
& $1.46 \pm 0.1$
& $1.36 \pm 0.09$
& $1.05 \pm 0.007$
& $0.647 \pm 0.044$
& $0.35 \pm 0.03$
& $0.184 \pm 0.012$
& $0.125 \pm 0.008$
\\
\hline
$\sigma_{H, {\rm th}}^{\eta; 2 \gamma}$ ($\mu{\rm b}$) 
& $1.49 \pm 0.10$
& $1.44 \pm 0.10$
& $1.40 \pm 0.10$
& $1.10 \pm 0.08$
& $0.72 \pm 0.05$
& $0.37 \pm 0.03$
& $0.21 \pm 0.01$ 
& $0.15 \pm 0.01$ 
\\
\hline
$R$ (in \%)
& $-$ 0.5
& $-$ 1.8
&   3.1
&   5.2
&  10.1
&   5.8
&  12.1
&  16.4 \\
\hline
\end{tabular}
\label{tab:2}
\end{center}

\begin{center}
  \caption{Comparison of the theoretical fit and data
    for the integral cross sections of the charge-exchange reaction
    $\pi^-  + p \to \eta'[\to 2 \gamma] + n$
     at beam momentum $P_{\rm Lab}$ running up to 40 GeV.}

\vspace*{.1cm}

\def\arraystretch{1.2}
\begin{tabular}{|c|c|c|c|c|c|c|}
\hline
\hline
$P_{\rm Lab}$ (GeV) & 15 & 20.2 & 25 & 30 & 39.1 & 40 \\
\hline
$\sigma_{H, {\rm data}}^{\eta'; 2 \gamma}$ ($\mu{\rm b}$)
& $ 0.19 \pm 0.02$
& $ 0.105 \pm 0.008$
& $ 0.081 \pm 0.006$
& $ 0.062 \pm 0.004$
& $ 0.042 \pm 0.004$
& $ 0.042 \pm 0.003$
\\
\hline
$\sigma_{H, {\rm th}}^{\eta'; 2 \gamma}$ ($\mu{\rm b}$) 
& $ 0.19 \pm 0.02$
& $ 0.12 \pm 0.01$
& $ 0.086 \pm 0.008$
& $ 0.065 \pm 0.006$
& $ 0.042 \pm 0.004$
& $ 0.041 \pm 0.004$
\\
\hline
$R$ (in \%) & 
0 & 1.1 & 0.8 & 4.1 & 1.1 & $-$ 2.5 \\
\hline
\hline
\end{tabular}
\label{tab:3}
\end{center}
\end{table}

\begin{table}[ht!] 
\begin{center}
  \caption{Comparison of the theoretical fit and data
    for the integral cross sections of the charge-exchange reaction
    $K^-  + p \to \bar K^0 + n$
     at beam momentum $P_{\rm Lab}$ running up to 40 GeV.}

\vspace*{.1cm}

\def\arraystretch{1.2}
\begin{tabular}{|c|c|c|c|c|c|c|c|c|}
\hline
\hline
$P_{\rm Lab}$ (GeV) & 5 & 7.7 & 8 & 14.3 & 25 & 34.6 & 40 \\
\hline
$\sigma_{H, {\rm data}}^{\bar K^0}$ ($\mu{\rm b}$)
& $ 151 \pm 10$
& $  71 \pm 13$
& $  78 \pm  7$
& $  28 \pm  6$
& $  13.9 \pm 1.1$
& $   7.5 \pm 1.1$
& $   6.8 \pm 0.5$
\\
\hline
$\sigma_{H, {\rm th}}^{\bar K^0}$ ($\mu{\rm b}$) 
& $ 134.61 \pm 18.57$
& $  74.48 \pm 10.27$
& $  70.60 \pm  9.74$
& $  30.89 \pm  4.26$
& $  13.69 \pm  1.89$
& $   8.48 \pm  1.17$
& $   6.84 \pm  0.94$
\\
\hline
$R$ (in \%) & 
-10.9 & 4.9 & - 9.5 & 10.3 & - 1.5 & 13.0 & 0.6\\
\hline
\end{tabular}
\label{tab:4}
\end{center}

\begin{center}
  \caption{Comparison of the theoretical fit and data
    for the integral cross sections of the charge-exchange reaction
    $\pi^-  + (A,Z) \to \pi^0 + (A,Z-1)$
     at beam momentum $P_{\rm Lab} = 40$ GeV.}

\vspace*{.1cm}

\def\arraystretch{1.2}
\begin{tabular}{|c|c|c|c|c|c|c|c|}
\hline
\hline
Nuclei &  H & Li & Be & C & Al & Fe & Cu \\
\hline
$\sigma_{Z, {\rm data}}^{\pi^0}$ ($\mu{\rm b}$)
& $ 10.52 \pm 0.5$
& $ 21.0 \pm 2.0$
& $ 23.4 \pm 2.5$
& $ 29.2 \pm 3.0$
& $ 48.0 \pm 5.0$
& $ 88.1 \pm 12.0$
& $ 95.0 \pm 12.0$
\\
\hline
$\sigma_{Z, {\rm th}}^{\pi^0}$ ($\mu{\rm b}$) 
& 10.16
& 19.52 
& 23.57
& 30.91 
& 52.38
& 84.31 
& 90.87 
\\
\hline
$R$ (in \%) & 
$-$ 3.4 & $-$ 7.1 & 7.1 & 5.9 & 9.1 & $-$ 4.3 & $-$ 4.3 \\
\hline\hline
\end{tabular}
\label{tab:5}
\end{center}

\begin{center}
  \caption{Comparison of the theoretical fit and data
    for the integral cross sections of the charge-exchange reaction
    $\pi^-  + (A,Z) \to \eta [\to 2 \gamma]  (A,Z-1)$
     at beam momentum $P_{\rm Lab} = 40$ GeV.}

\vspace*{.1cm}

\def\arraystretch{1.2}
\begin{tabular}{|c|c|c|c|c|c||}
\hline
\hline
Nuclei & H & C & Al & Fe & Cu \\
\hline
$\sigma_{Z, {\rm data}}^{\eta  \to  2 \gamma}$ ($\mu{\rm b}$)
& $ 1.46 \pm 0.1$
& $ 4.0 \pm 1.0$
& $ 6.7 \pm 2.0$
& $ 12.0 \pm 3.6$
& $ 13.0 \pm 4.0$
\\
\hline
$\sigma_{Z, {\rm th}}^{\eta  \to  2 \gamma}$ ($\mu{\rm b}$) 
& 1.43 
& 4.36 
& 7.39
& 11.90
& 12.83 
\\
\hline
$R$ (in \%) & 
$-$ 1.8 & 9.1 & 10.3 & $-$ 0.8 & $-$ 1.3 \\
\hline\hline
\end{tabular}
\label{tab:6}
\end{center}
\end{table}

\section{Conclusions}

We presented the high-quality fit of the integral cross sections of the CEX reactions
with pion and kaon beams in wide region of the $P_{\rm Lab}$ running from 5 Gev and
up to 200 GeV proton target and at different nuclei. 
For the first time we described data with approximate errors of $\simeq 10\%$.
Our results can be used as the data set for analysis of future experimental studies
of the CEX cross sections. 
For the interval $P_{\rm Lab} = 5-50$ GeV our results are in particular important
for the sensitivity estimate of the ongoing and planned searches in the NA64h experiment
at CERN  which used the CEX reactions with charged pion and kaon beams as a source
of light neutral mesons. 

\begin{acknowledgments} 

We would like to thank our NA64 colleagues for useful discussions.   
This work was funded by FONDECYT (Chile) under Grants
No. 1230160 and No. 1240066, and
by ANID$-$Millen\-nium Program$-$ICN2019\_044 (Chile).

\end{acknowledgments}

\newpage

\begin{thebibliography}{999}

\bibitem{NA64h:2024mah}
Yu.~M.~Andreev \textit{et al.} (NA64 Collaboration),
Phys. Rev. Lett. \textbf{133}, 121803 (2024). 

\bibitem{Gninenko:2023rbf}
S.~N.~Gninenko, D.~V.~Kirpichnikov, S.~Kuleshov, V.~E.~Lyubovitskij,
and A.~S.~Zhevlakov,
Phys. Rev. D \textbf{109}, 075021 (2024). 

\bibitem{Gninenko:2014sxa}
S.~N.~Gninenko,
Phys. Rev. D \textbf{91}, 015004 (2015).   

\bibitem{Gninenko:2015mea}
S.~N.~Gninenko and N.~V.~Krasnikov,
Phys. Rev. D \textbf{92}, 034009 (2015);  
Mod. Phys. Lett. A \textbf{31}, 1650142 (2016).

\bibitem{Gninenko:2024iyx}
S.~N.~Gninenko, D.~V.~Kirpichnikov, N.~V.~Krasnikov, S.~Kuleshov, V.~E.~Lyubovitskij, 
and A.~S.~Zhevlakov,
JHEP \textbf{06}, 235 (2025). 

\bibitem{Gninenko:2024ujp}
S.~N.~Gninenko, N.~V.~Krasnikov and V.~A.~Matveev,
Natural Sci. Rev. \textbf{1}, 5 (2024), 

\bibitem{Gninenko:2025xmb}
S.~N.~Gninenko and N.~V.~Krasnikov,
Phys. Rev. D \textbf{111}, 11 (2025). 

\bibitem{Barnes:1976ek}
A.~V.~Barnes \textit{et al.}, 
Phys. Rev. Lett. \textbf{37}, 76 (1976);
O.~I.~Dahl \textit{et al.}, 
Phys. Rev. Lett. \textbf{37}, 80 (1976);
W.~D.~Apel \textit{et al.} (Serpukhov-CERN Collaboration),
Phys. Lett. B \textbf{83}, 131 (1979); 
Nucl. Phys. B \textbf{152}, 1 (1979);  
Nucl. Phys. B \textbf{154}, 189 (1979); 
V.~I.~Belousov \textit{et al.}, 
Phys. Lett. B \textbf{43}, 76 (1973);
V.~D.~Apokin \textit{et al.}, 
Sov. J. Nucl. Phys. \textbf{46}, 644 (1987); 
S.~V.~Donskov \textit{et al.},  
Eur. Phys. J. C \textbf{73}, 2614 (2013).

\bibitem{Apokin:1981iv}
V.~D.~Apokin \textit{et al.}, 
Sov. J. Nucl. Phys. \textbf{35}, 219 (1982). 

\bibitem{Astbury:1966ufw}
P.~Astbury \textit{et al.},  
Phys. Lett. \textbf{23}, 396 (1966);  
G.~W.~Brandenburg \textit{et al.},  
Phys. Rev. D \textbf{15}, 617 (1977); 
W.~D.~Apel \textit{et al.} (Serpukhov-CERN Collaboration),
Nucl. Phys. B \textbf{129}, 275 (1977); 
F.~G.~Binon \textit{et al.} (Serpukhov-Brussels-Annecy(LAPP)
and Serpukhov-CERN Collaborations),
Sov. J. Nucl. Phys. \textbf{33}, 542 (1981);
Nuovo Cim. A \textbf{64}, 89 (1981).

\bibitem{Strakovsky:2016azh}
I.~I.~Strakovsky,
EPJ Web Conf. \textbf{130}, 06005 (2016); 
M.~Amaryan \textit{et al.} (KLF Collaboration),
[arXiv:2008.08215 [nucl-ex]];
M.~J.~Amaryan \textit{et al.} (KLF Collaboration),
Mod. Phys. Lett. A \textbf{39}, 2450063 (2024). 

\bibitem{Regge}
T.~Regge,
Nuovo Cim. \textbf{14}, 951 (1959); 
V.~N.~Gribov,
Sov. Phys. JETP \textbf{26}, 414(1968); 
P.~D.~B.~Collins,
{\it An Introduction to Regge Theory and High-Energy Physics}, 
Cambridge Univ. Press, 2009.

\bibitem{Nys:2018vck}
J.~Nys \textit{et al.} (JPAC Collaboration),
Phys. Rev. D \textbf{98}, 034020 (2018).

\end{thebibliography}
\end{document}